\documentclass{article}

    \usepackage[preprint, nonatbib]{neurips_2020}

\usepackage[utf8]{inputenc} 
\usepackage[T1]{fontenc}    
\usepackage{hyperref}       
\usepackage{url}            
\usepackage{booktabs}       
\usepackage{amsfonts}       
\usepackage{nicefrac}       
\usepackage{microtype}      
\usepackage{graphicx}

\title{Human computation requires and enables a new approach to ethical review}

\author{
  Libu\v{s}e H.~Vep\v{r}ek\thanks{Ludwig-Maximilians-Universität München}\\
  Human Computation Institute\\
  Ithaca, NY 14850 \\
  \texttt{l.veprek@lmu.de} \\
  \And
  Patricia Seymour \\
  Human Computation Institute\\
  Ithaca, NY 14850 \\
  \texttt{pannseymour@yahoo.com} \\
  \And
  Pietro Michelucci \\
  Human Computation Institute\\
  Ithaca, NY 14850 \\
  \texttt{pem@humancomputation.org} \\
}

\begin{document}

Note: This is a pre-publication draft submitted to 34th Conference on Neural Information Processing Systems (NeurIPS 2020), Vancouver, Canada.

\maketitle

\begin{abstract}
With humans increasingly serving as computational elements in distributed information processing systems and in consideration of the profit-driven motives and potential inequities that might accompany the emerging thinking economy[1], we recognize the need for establishing a set of related ethics to ensure the fair treatment and wellbeing of online cognitive laborers and the conscientious use of the capabilities to which they contribute. Toward this end, we first describe human-in-the-loop computing in context of the new concerns it raises that are not addressed by traditional ethical research standards. We then describe shortcomings in the traditional approach to ethical review and introduce a dynamic approach for sustaining an ethical framework that can continue to evolve within the rapidly shifting context of disruptive new technologies.
\end{abstract}

\section{Introduction}
A new branch of artificial intelligence called ``human computation'' has emerged over the last 15 years that combines the respective strengths of humans and computers to tackle problems that cannot be solved in other ways[2]. Information processing systems based on this approach often employ online crowdsourcing to delegate to humans cognitive ``microtasks'' that elude the capabilities of machine-based methods. Real-world human computation systems are already advancing cancer[3], HIV[4], and Alzheimer's[5] research, diagnosing malaria[6] in sub-Saharan Africa, reducing female genital mutilation in Tanzania[7], predicting flood effects in Togo[8], endowing the blind with real-time scene understanding[9], expediting disaster relief despite language barriers and failing infrastructure[10], rewriting our understanding of cosmology[11][12] and improving predictions in conservation science[13].

Four main paradigms exist for engaging people in human computation tasks. One of the first online human computation systems, \emph{reCAPTCHA}, operates on a quid-pro-quo basis, requiring a person to digitize distorted text to demonstrate being human, which provides access to a website while simultaneously contributing to a massive text digitization project[14]. Modern versions of this, such as \emph{hCaptcha}, require people to select all images from a provided set that contain some class of objects, such as traffic lights. This generates revenue by providing a data-labelling service that is used to help train customers' machine learning models[15]. Another paradigm, citizen science, entices public volunteers to participate in the scientific process[16] through which they collect and analyze research data in exchange for hands-on opportunities to learn about various research topics. Citizen science has been steadily gaining popularity via community connectors like \emph{SciStarter.org} and curation platforms like \emph{Zooniverse}. In a third engagement paradigm, ``clickworkers'' are typically paid via crowdsourcing marketplaces, such as \emph{Amazon Mechanical Turk} and \emph{ClickWorker.com} to participate in various online tasks that might be used for science, marketing, or product development applications.[17] \emph{Mooqita} represents a version of paid-crowdsourcing that uses massive open online courses (MOOC) as a way to onboard new employees for salaried jobs involving online cognitive labor for a single organization rather than a marketplace. In the final and most recent paradigm, popular online games such as \emph{EVE Online} and \emph{Borderlands Science} have begun embedding microtasks within existing gameplay to engage potentially millions of online gamers who opt-in to participate in exchange for various advancement opportunities in the games[18].

Humans are squarely involved in human computation systems, and although most configurations do not constitute human subjects research, they often pose new ethical dilemmas. Thus, traditional research values and review practices typically do not apply to human computation and seem inadequate to address the many new human-centered contexts produced by this growing field of study.

This becomes especially apparent in ethical review processes. Currently, online Citizen Science research is reviewed by the same standards as clinical trials although the review board's understanding of ``human subject'' or ``research participant'' does not fit the role of participants in online citizen science. Citizen scientists are not mere ``human subjects'' in a research study, but can be researchers, human subjects, or sometimes both at the same time. This implies that their role has to be communicated and elucidated to the traditional review boards. These new role allocations also come along with different ideas and expectations for the different stakeholders, especially those performing tasks. For example, David Resnik shows that if the role of citizen scientists exceeds being a mere human subject, they may feel more connected and hence more ownership over their collected data[19]. Moreover, in the arena of citizen science and human computation, traditional IRB is not clearly mandated and, when used, it rigidly enforces ethical standards most often associated with biomedical research.

Ironically, human computation provides a potential solution to these ethical challenges. Herein we argue for a new approach to ethical oversight that addresses the needs of online citizen science and the new forms of human-computer collaboration in a digital age. First we explain how a distinction between morals and ethics can be useful for this endeavor. Based on this distinction we explore how we might appeal to the participatory methods of human computation to design a technosocial platform that enables the curation of a ``living'' set of ethics and crowdsources the application of those ethics to a suitable review process.

\section{Morality vs. Ethics}
In everyday life ``morality'' and ``ethics'' are used synonymously to describe the ``good'' in contrast to the ``bad'' or to distinguish between ``right'' and ``wrong''.  Formally, however, Ethics is considered a branch of philosophy that studies morality[20], or in the words of the German philosopher Dietmar H{\"u}bner: “Morality is the object, ethics is science.”[21] This brings us to the question of: what ismorality? According to H{\"u}bner, then is ``a system of norms, whose subject is human behavior and which claims unconditional validity''(ibid.: 13, transl. b.t.a.). This means that different ``morals'' exist in different contexts such as different cultures or political currents, and there even exist different morals for specific groups of people like doctors, journalists and scientists. To decide whether a given moral is right or wrong therefore requires reflecting on the normative context in which it occurs. Thus, the aim of differentiating between ``morality'' and ``ethics'' is not to enforce correct usage, but about orienting our perspective as we consider ethical review in the new kinds of digital collaborations that manifest in human computation systems such as online citizen science. 

We can now relate to two different layers that need to be considered: The first layer consists of the moral framework in the field. We need to understand what the moral values of the different stakeholders in human computation and citizen science are: For example, what is important to citizen science participants?  What do they expect from their participation and how do they want to be acknowledged? This understanding of “moral framework” could be related to what Lisa Rasmussen calls the “citizen science ethos”[23]. To identify moral values in online citizen science, we introduced an online discussion forum and invited both citizen scientists and citizen science practitioners to contribute [22]. The analysis of the different discussion thread shows that mutual respect, inclusion and transparency a.o. belong to the ``moral framework'' of citizen science. This could then inform the second layer which consists of reflecting this moral framework in a set of ethical guidelines, which could both inform the design and execution of citizen science projects as well as evaluation by ethical oversight committees, who can then review citizen science research in accordance with community values. Of course, these ethical guidelines would need to remain flexible as we may discover that values that apply to most citizen science projects may still be ill-suited to certain specific studies.

\section{Traditional IRB}
The difference between our attitudes toward people who participate in research can perhaps be seen best in the debate between using the term ``participant'' or ``subject'' to refer to someone who volunteers their body, mind, and time to research. In 2014, Public Responsibility in Medicine and Research (PRIM\&R) put forth the following statement on this debate: ``subject'' is the most appropriate title for those involved in research studies (recognizing, however, that in some instances ``participant'' may be appropriate; for example, in community-based participatory research). In the world of citizen science, however, participant activities may be more closely aligned with the work that scientists do. This means that currently there is no existing model of review that appropriately assigns autonomy to those engaged. 

Moreover, the absence of ethical guidelines for citizen science creates a dilemma for independent/institutional reviews (IRB) or ethics review boards (ERB) because they have to choose between either applying ethical guidelines that do not fit the application or making their own decisions about what’s right or wrong. The lack of consistent interpretation of research studies has prompted the National Institutes of Health in the United States to mandate a single IRB review for some federally funded research. The researcher, at the time of the grant application, identifies an IRB that will serve as the only IRB for the project. This removes the multitude of review decisions, consent variation and timing delays that routinely plague even the most prestigious and well-funded research institutions.\footnote{Throwing technology at the issue of IRB review was thought to be an answer to better organize and streamline what had, for many years, been a paper-based process.  The result of the development and use of these electronic, online platforms has not been particularly helpful because it applies technology to an inefficient process.  To make matters worse, the technology is customized to the already-broken system, perpetuating the pre-existing problems in an online context. Moreover, the systems available in the U.S. do not communicate between institutions and do not solve the problem of inconsistent decisions and documents.}

Lastly, we build on our own experiences with IRB processes in the field of human computation-based citizen science: Working with a traditional IRB highlighted the inconsistencies and infeasibility of using the traditional US biomedical oriented approach of most IRBs. Fortunately, the experience was mitigated by a knowledgeable and flexible liaison who helped to inform the members of the IRB of the differences in human computation from biomedical research. For example, the concept of risk and benefit, as defined by the U.S. regulations and guidance did not apply in a straightforward manner to human computation. The protectionism of the US regulations may not be applicable to human computation nor does it apply to the participant's autonomy when performing tasks in this arena. Each difference in approach and regulatory necessity had to be addressed, thus causing delays and redundancies that was not conducive to efficient review.

Why should citizen science, and indeed the broader space of human computation, as a relatively new approach to research, enter the fray of poorly organized and inefficient traditional review processes just because review has always been done this way? Instead, we envision a triage-based approach that determines whether traditional research ethics review applies to a research project based on the project goals and participant task requirements and flips four key aspects of traditional IRB. It also involves input from members of the human computation community at large.

\section{Flipping IRB}
Traditionally, IRB has often been compulsory and seemed like an adversarial process, where reviewers were there to poke holes in the researcher's ethical approach and eventually return a judgment. This process could be  recast as a collaborative one in which the role of the IRB expert is to help researchers align their methods with established ethical guidelines toward achieving their research goals.

Traditionally, an application is filled out cautiously and word choices are carefully made to elicit an approval outcome. The application then disappears into a black box called the IRB process, where mystical and sometimes random-seeming things happen until a determination comes out the other end. We believe full transparency is critical for building trust and working toward a shared goal. Once aligned with an IRB expert who can direct the ethical review of a project as a collaborator, there is a clear path to approval.

Traditionally, applications are isolated documents that are filed away and never seen again (until it's time to amend or renew). But that means that even if two projects by different researchers are very similar, they each have to go through the same time-intensive, expensive, and laborious process. We think that if your project is like someone else's then we only need to consider the parts that are different. We suggest that if we begin building a protocol repository, we can assist researchers to use designs and approaches that have already been approved thus decreasing the issues that have to be reviewed time and time again.

Traditionally, there was a ready roster of mostly the same people who were on tap to review applications as needed. We think the only time you need a panel is when there is something that isn't clear cut, and in that case, let's enlist our community of peers, including human computation community volunteers who have agreed to be on call.

\section{A human computation approach to ethical review}
Today, we have an opportunity to build a technosocial platform that turns this vision into reality. We apply our own human computation approaches to building consensus and determining the right division of labor between humans and machines. We also draw on our experience related to creating a human computation platform that continues to engage over 30,000 volunteers to motivate community participation in the review process. Indeed, we have begun to build this - it is called Civium, and its main purpose is to make human computation research and applications more transparent, trustworthy, and sustainable.\footnote{An effort based on a similar logic is \emph{OpenReview}, a web interface with underlying database API that aims at advancing openness in the scientific peer review process. We would like to thank the anonymous reviewer for drawing our attention to this analogy.}

For starters, we created an experimentation toolkit that makes it possible to clone a citizen science project with a single click to create a sandbox version for running experiments without affecting the live platform or data quality. This sandbox environment can be used to run an experiment that studies the behaviors of the citizen science volunteers. In this case, volunteers are not working alongside scientists analyzing data, they are being studied by scientists to help improve our understanding about how to design effective citizen science platforms. For example, in one case study we investigated human/AI-partnerships in the online game \emph{Stall Catchers}. By including surveys in different stages of the experiment and analyzing the collected (meta) data we could gain insights into when and why users trusted the AI-assistant and when they questioned its' skill. This configuration, however, goes beyond traditional citizen science, might suggest the need for ethical review.

\begin{figure}[h]
  \centering
  \includegraphics[width=1\linewidth]{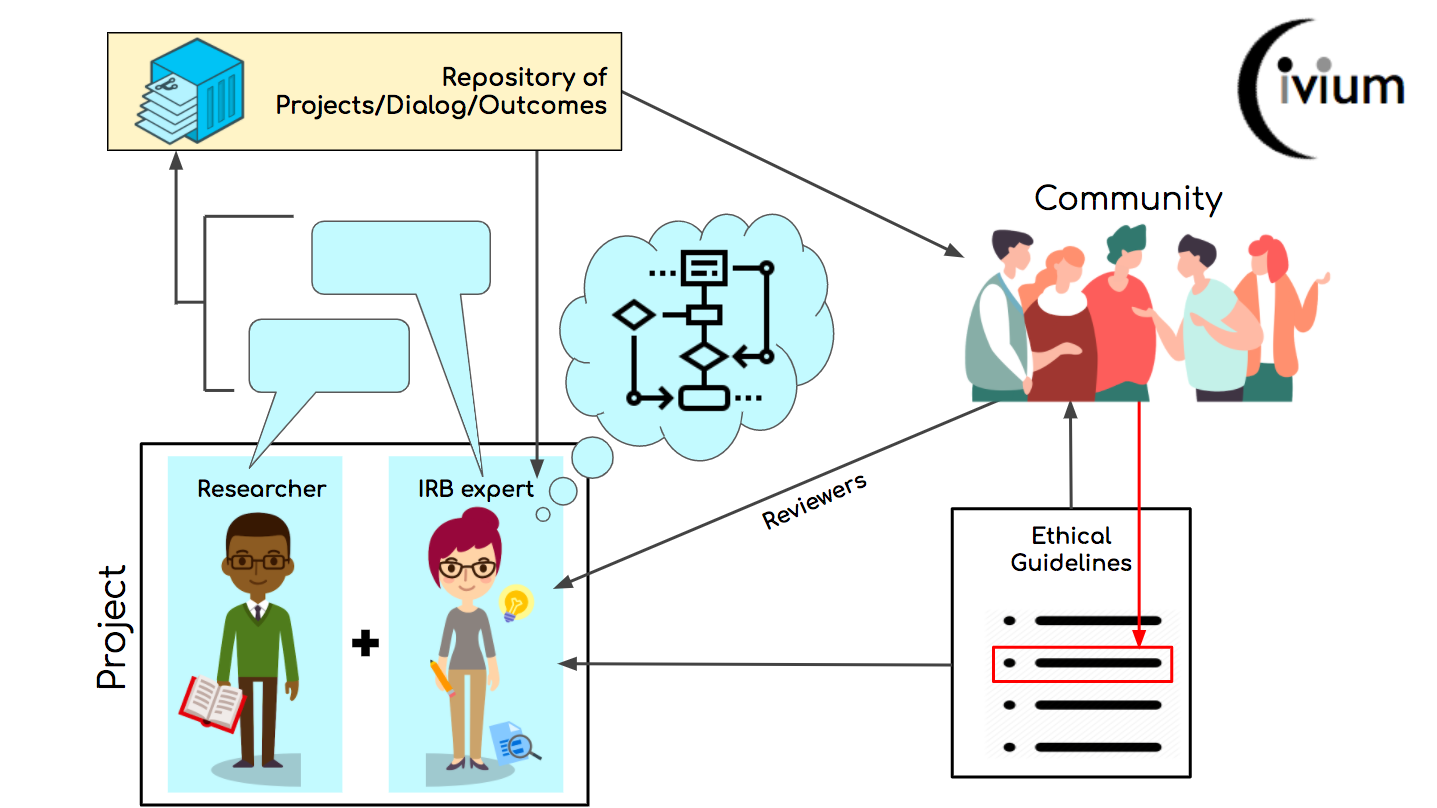}
  \caption{Activity diagram of technosocial platform for ethical review}
\end{figure} 

To make that process quick, informative, and painless, we are integrating a new IRB process (see Figure 1) into the Civium environment. That means that instead of packaging up a description of an experiment for the IRB expert, the expert can enter into the sandbox with the researcher and examine the experiment to understand its design (see Figure 2 and 3 in the appendix) and see from the standpoint of a participant how it will actually work (see Figure 4 in the appendix). The IRB expert can make comments and suggest edits directly on the interface in support of aligning the research goals with the community's current ethical standards.This new procedure should minimize the time a reviewer has to work on a proposal, since the submitted proposal could entail links to the sandbox experiment allowing the reviewer to gain a better understanding of the design logic and the interface. Traditionally, the reviewer would have to think about how the experiment would probably work based on written material and some screenshots. The explicit role of the IRB expert is to shepherd the review process and activate various assistive mechanisms as needed in service of that goal.

Meanwhile, the system logs the issues that arise and the ensuing dialog around those issues, as well as any implemented solutions. These are made transparent and accessible in a repository that connects these to a snapshot of the sandbox itself. This introduces something completely new and powerful for ethical review - something we demand for our legal system and something we aim for in scientific research, but that is missing in IRB, which is reproducibility.
And when questions arise that are not addressed or that are ambiguous under the current set of ethical guidelines, the IRB expert invokes members of the community to review the issues. This is analogous to a section editor for an academic journal finding reviewers for a manuscript. And the platform can help manage the recruitment of these community contributors.

Finally, and critically, the outcome of this process can not only resolve the ethical dilemma for the researchers, it can inform amendments to the ethical guidelines themselves, so that they can continue to grow with our understanding of human-in-the-loop computing and fit the needs and circumstances of this growing community.

\section*{Broader Impact}
Herein we address the need for a general-purpose mechanism for AI governance that can evolve with our understanding of the field. Not only does AI raise issues of autonomy, labor, and equitability, but with increased reliance on systems that employ human cognition, the ethical waters become even murkier. Our approach seeks to crowdsource the evaluation of the risks and rewards of situated AI systems as well as human-computer collaborations, and aims at seeding and curating a set of related ethics to ensure the fair treatment and wellbeing of humans in-the-loop. To ensure that the needs of all stakeholders are taken into account we include diverse perspectives in a maximally transparent process. For example, all reports will be stored in a searchable public repository. However, despite our best intentions and planning, there are always the “unknown unknowns” that can arise when a platform like this goes live. To address these risks, we will remain vigilant to the system’s behavior and leverage community monitoring and feedback loops intrinsic to our approach.

\begin{ack}
We would like to thank Eglė Marija Ramanauskaitė for her great assistance and preparation of the activity diagram and Percy Mamedy for his work on the implementation of the platform. We also wish to show our appreciation to all participants of our discussion on reinventing IRB as well as the contributors in the citizen science forum for their helpful insights and the fruitful discussions.

Libuše Hannah Vepřek, Patricia Seymour and Pietro Michelucci declare that they have no conflict of interest.
\end{ack}

\section*{References}
\medskip
\small
[1]	P. Michelucci, ``How do we create a sustainable thinking economy?," {\it Toward Data Science}, Nov. 04, 2019. https://towardsdatascience.com/how-do-we-create-a-sustainable-thinking-economy-4d77839b031e (accessed Feb. 04, 2020).

[2]	P. Michelucci and J. L. Dickinson, ``The power of crowds," {\it Science}, vol. 351, no. 6268, pp. 32–33, Jan. 2016, doi: 10.1126/science.aad6499.

[3]	F. J. Candido Dos Reis et al., ``Crowdsourcing the General Public for Large Scale Molecular Pathology Studies in Cancer," {\it EBioMedicine}, vol. 2, no. 7, pp. 681–689, Jul. 2015, doi: 10.1016/j.ebiom.2015.05.009.

[4]	F. Khatib et al., ``Crystal structure of a monomeric retroviral protease solved by protein folding game players," {\it Nat. Struct. Mol. Biol.}, vol. 18, no. 10, pp. 1175–1177, Sep. 2011, doi: 10.1038/nsmb.2119.

[5]	O. Bracko et al., ``High fat diet worsens pathology and impairment in an Alzheimer's mouse model, but not by synergistically decreasing cerebral blood flow," {\it bioRxiv}, p. 2019.12.16.878397, Dec. 2019, doi: 10.1101/2019.12.16.878397.

[6]	M. A. Luengo-Oroz, A. Arranz, and J. Frean, ``Crowdsourcing Malaria Parasite Quantification: An Online Game for Analyzing Images of Infected Thick Blood Smears," {\it J. Med. Internet Res.}, vol. 14, no. 6, Nov. 2012, doi: 10.2196/jmir.2338.

[7]	M. O. Leaders, ``Put Rural Tanzania on the Map," {\it Read, Write, Participate}, Apr. 19, 2018. https://medium.com/read-write-participate/put-rural-tanzania-on-the-map-79d0888df210 (accessed May 02, 2019).

[8]	P. Suarez, ``Rethinking Engagement: Innovations in How Humanitarians Explore Geoinformation," {\it ISPRS Int. J. Geo-Inf.}, vol. 4, no. 3, pp. 1729–1749, Sep. 2015, doi: 10.3390/ijgi4031729.

[9]	J. P. Bigham et al., ``VizWiz: Nearly Real-time Answers to Visual Questions," in {\it Proceedings of the 23Nd Annual ACM Symposium on User Interface Software and Technology}, New York, NY, USA, 2010, pp. 333–342, doi: 10.1145/1866029.1866080.

[10]	P. Meier, ``Human Computation for Disaster Response," in {\it Handbook of Human Computation}, P. Michelucci, Ed. New York, NY: Springer, 2013, pp. 95–104.

[11]	A. J. Westphal et al., ``Evidence for interstellar origin of seven dust particles collected by the Stardust spacecraft," {\it Science}, vol. 345, no. 6198, p. 786, Aug. 2014, doi: 10.1126/science.1252496.

[12]	C. N. Cardamone et al., ``Galaxy Zoo Green Peas: Discovery of A Class of Compact Extremely Star-Forming Galaxies," {\it Mon. Not. R. Astron. Soc.}, vol. 399, no. 3, pp. 1191–1205, Nov. 2009, doi: 10.1111/j.1365-2966.2009.15383.x.

[13]	S. Kelling et al., ``eBird: A Human/Computer Learning Network for Biodiversity Conservation and Research," presented at the {\it Twenty-Fourth IAAI Conference}, Jul. 2012, Accessed: May 02, 2019. [Online]. Available: https://www.aaai.org/ocs/index.php/IAAI/IAAI-12/paper/view/4880.

[14]	L. von Ahn, B. Maurer, C. McMillen, D. Abraham, and M. Blum, ``reCAPTCHA: Human-Based Character Recognition via Web Security Measures," {\it Science}, vol. 321, no. 5895, pp. 1465–1468, Sep. 2008, doi: 10.1126/science.1160379.

[15]	``Moving from reCAPTCHA to hCaptcha," {\it The Cloudflare Blog}, Apr. 08, 2020. https://blog.cloudflare.com/moving-from-recaptcha-to-hcaptcha/ (accessed Oct. 09, 2020).

[16]	F. Heigl, B. Kieslinger, K. T. Paul, J. Uhlik, and D. D\"{o}rler, ``Opinion: Toward an international definition of citizen science," {\it Proc. Natl. Acad. Sci.}, vol. 116, no. 17, p. 8089, Apr. 2019, doi: 10.1073/pnas.1903393116.

[17]	J. Chandler, G. Paolacci, and P. Mueller, ``Risks and Rewards of Crowdsourcing Marketplaces," in {\it Handbook of Human Computation}, P. Michelucci, Ed. New York, NY: Springer, 2013, pp. 377–392.

[18]	J. Waldisp\"{u}hl, A. Szantner, R. Knight, S. Caisse, and R. Pitchford, ``Leveling up citizen science," {\it Nat. Biotechnol.}, vol. 38, no. 10, pp. 1124–1126, Oct. 2020, doi: 10.1038/s41587-020-0694-x.

[19]	D. B. Resnik, ``Citizen Scientists as Human Subjects: Ethical Issues," {\it Citiz. Sci. Theory Pract.}, vol. 4, no. 1, Art. no. 1, Mar. 2019, doi: 10.5334/cstp.150.

[20]	J. Deigh, {\it An Introduction to Ethics}. Cambridge: Cambridge University Press, 2010.

[21]	D. H\"{u}bner, Einf\"{u}hrung in {\it die philosophische Ethik, 2nd ed.} G\"{o}ttingen: UTB GmbH, 2018.

[22] 	``Citizen Science Ethics" {\it Human Computation Institute forum}, 2020. https://forum.hcinst.org/c/citsci-ethics/13 (accessed Nov. 18, 2020).

[23]    L. M. Rasmussen, “Confronting Research Misconduct in Citizen Science,"Citiz. Sci. Theory Pract., vol. 4,no. 1, Art. no. 1, Mar. 2019, doi: 10.5334/cstp.207.

\appendix
\section*{Appendix}

\begin{figure}[h]
  \centering
  \includegraphics[width=1\linewidth]{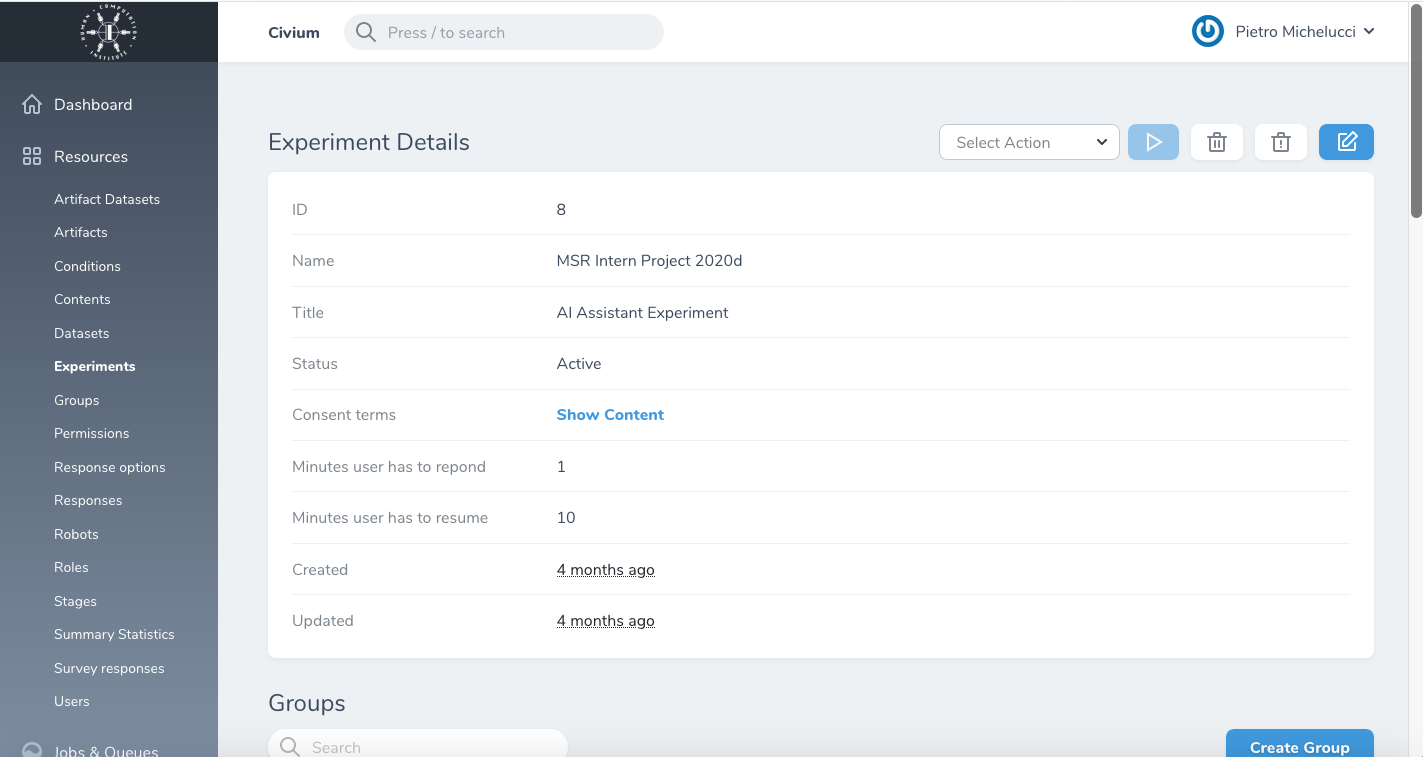}
  \caption{Screenshot of nova backend: example experimental design (prototype)}
\end{figure}

\begin{figure}[h]
  \centering
  \includegraphics[width=1\linewidth]{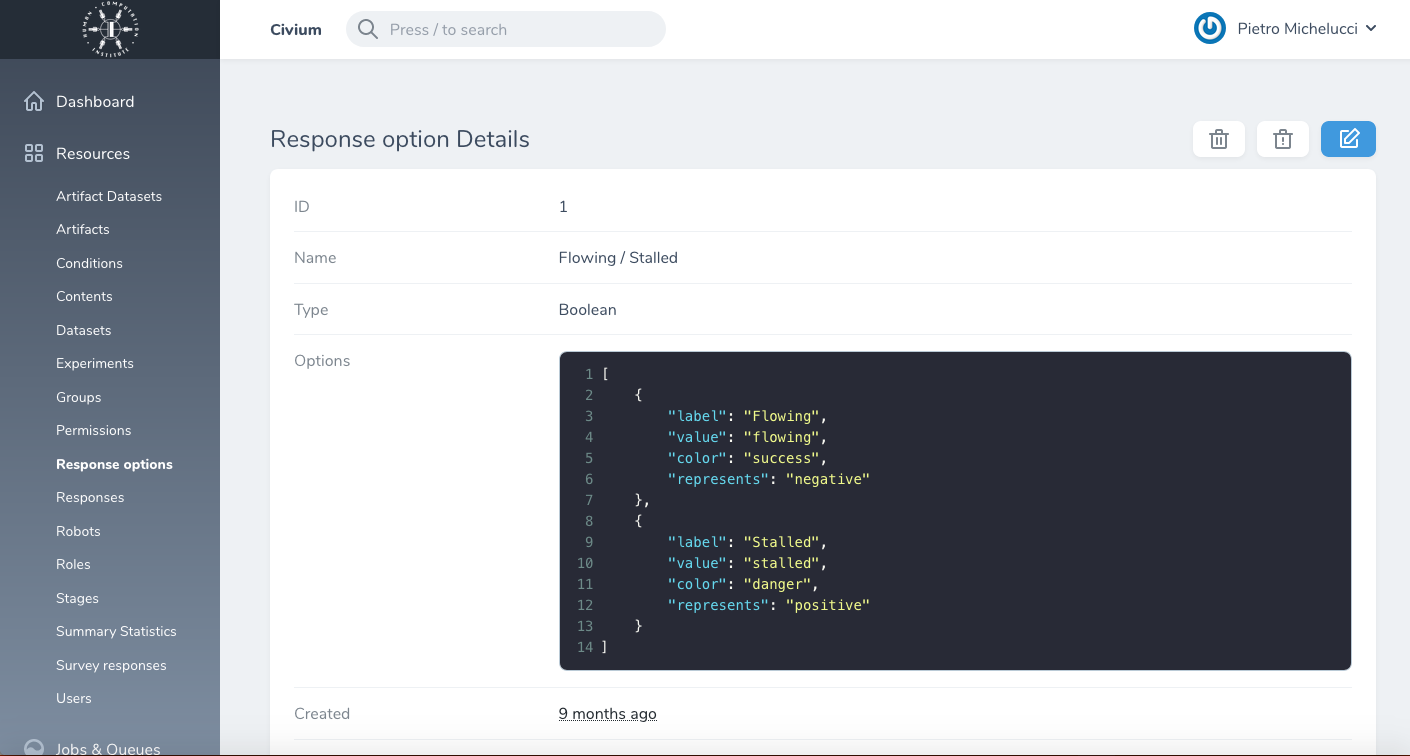}
  \caption{Screenshot of nova backend: example experimental design (prototype)}
\end{figure}

\begin{figure}[h]
  \centering
  \includegraphics[width=1\linewidth]{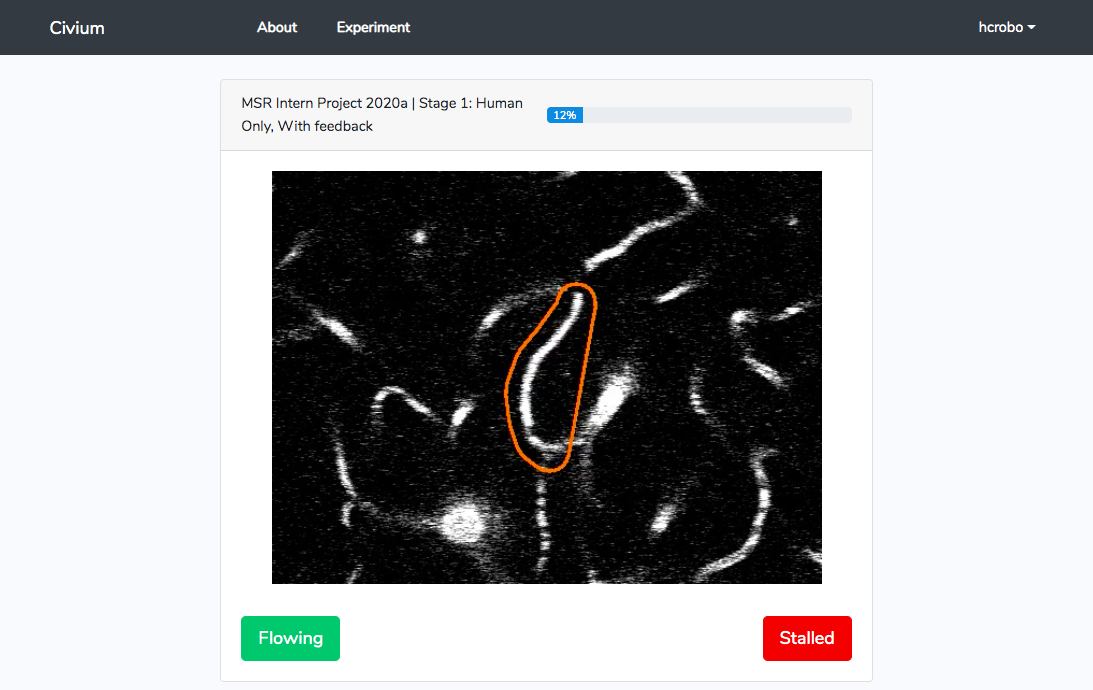}
  \caption{Screenshot of example user interface for experiments (prototype)}
\end{figure}

\end{document}